\documentclass[
  aps,  
  prb,
  twocolumn,
  longbibliography,
  superscriptaddress,
  preprintnumbers,
  floatfix,
]{revtex4-2}

\usepackage{graphicx}
\usepackage{mathtools}
\usepackage{dcolumn}
\usepackage{bm}
\usepackage{hyperref}
\renewcommand\vec\bm  
\DeclareMathOperator\sign{sign}
\usepackage{braket}
\usepackage[caption=false]{subfig}
\usepackage{microtype}
\usepackage[acronym]{glossaries}
\usepackage{hyperref}
\usepackage{cleveref}
\usepackage{amsfonts}
\usepackage{xcolor}

\newacronym{ll}{LL}{Landau level}

\begin{document}

\title{Unconventional thermoelectric transport in  tilted  Weyl semimetals 
}

\author{Thorvald M. Ballestad}
\affiliation{%
  Center for Quantum Spintronics, Department of Physics, Norwegian University of Science and Technology, Trondheim, Norway
}%

\author{Alberto Cortijo}%
\affiliation{%
 Departamento de Física de la Materia Condensada, Universidad Autónoma de Madrid, Madrid E-28049, Spain}

\author{Mar\'ia A. H. Vozmediano}%
\affiliation{%
Instituto de Ciencia de Materiales de Madrid,
CSIC, Cantoblanco, E-28049 Madrid, Spain}

\author{Alireza Qaiumzadeh}%
\affiliation{%
  Center for Quantum Spintronics, Department of Physics, Norwegian University of Science and Technology, Trondheim, Norway
}%

\date{\today}

\begin{abstract}
  We analyze the effect of the tilt on the transverse  thermoelectric coefficient of Weyl semimetals in the  \emph{conformal} limit, i.e., zero temperature and zero chemical potential. Using the Kubo formalism, we find a nonmonotonic behavior of the thermoelectric conductivity as a function of the tilt perpendicular to the magnetic field, and a linear behavior when the tilt is aligned to the magnetic field. An ``axial Nernst" current (chiral currents counter propagating perpendicularly to the magnetic field and the temperature gradient) is generated in inversion symmetric Weyl materials when the tilt vector has a projection in the direction of the magnetic field. This analysis will help in the design and interpretation of thermoelectric transport experiments in recently discovered topological quantum materials.
\end{abstract}

\maketitle
\section{Introduction}
Emergent low-energy massless quasiparticles in three-dimensional (3D) Dirac and Weyl semimetals (WSMs) provide a testbed for investigation of interesting phenomena  beyond standard relativistic quantum field theory (QFT) and Landau Fermi liquid paradigm \cite{RevModPhys.93.025002,armitageWeylDiracSemimetals2018,2022SciA....8.1076L}. 
As opposed to the case in relativistic QFT,  material models are not Lorentz invariant. In addition to the fact that massless Weyl quasiparticles move at the Fermi velocity - a hundred times smaller than the speed of light - other Lorentz breaking terms may appear in the low energy description of Dirac materials \cite{Kost22}.
One generic feature of  the dispersion relation in most 3D Dirac and WSMs not studied in QFT, is the tilt of the cones \cite{PhysRevLett.129.056601}. According to its value,  topological quantum semimetals are classified into 
 type-I,  having  pointlike Fermi surface with vanishing density of states, and  type-II WSMs with overtilted cones that have a hyperbolic  Fermi surface with finite density of states at the Fermi level~\cite{RevModPhys.93.025002,soluyanovTypeIIWeylSemimetals2015,PhysRevLett.115.265304}. In some systems the tilting angle of Weyl cones and the topological Lifshitz phase transition \cite{volovikTopologicalLifshitzTransitions2017} between type-I and type-II Weyl phases can be controlled by external perturbations \cite{PhysRevB.94.121106,FelserLif19}.

One of the most interesting aspects of the physics of massless Dirac fermions in QFT is related to quantum anomalies~\cite{bertlmannAnomaliesQuantumField1996}.
They occur when a  symmetry of a classical field theory action does not survive quantization  \cite{Marsh,zeeQuantumFieldTheory2010,Fradkin,Shifman}.
The best known quantum anomaly is the chiral anomaly associated to the non-conservation of the independent number of left- and right-handed Weyl fermions in a 3D massless Weyl system~\cite{Marsh, zeeQuantumFieldTheory2010}. First identified in QFT as responsible for the fast decay of the neutral pion into two photons~\cite{adlerAxialVectorVertexSpinor1969,bellPCACPuzzleP01969}, is living a new bloom  after its experimental realization in the modern condensed matter systems~\cite{Ong2021} where
 it gives rise to exotic thermo- and  magnetoelectric transport phenomena ~\cite{burkovChiralAnomalyTransport2015, wehlingDiracMaterials2014, burkovTopologicalSemimetals2016,PhysRevB.89.085126,PhysRevResearch.2.013088,PhysRevResearch.2.033511,PhysRevB.100.085406}.

A lesser known quantum anomaly in condensed matter physics is the  scale or conformal anomaly  
 associated  the quantum breakdown of scale invariance. The conservation of the corresponding Noether current (dilatation current) implies  the traceless of the energy-momentum tensor~\cite{zeeQuantumFieldTheory2010,Fradkin,Shifman}, a condition  that is violated after quantization in some anomalous QFTs. 
In the gravity context, 
it was recently shown that the scale anomaly leads to a new electromagnetic transport  effect in the presence of an inhomogeneous gravitational background, the scale magnetic effect~\cite{chernodubAnomalousTransportDue2016}. More recently,  using the Luttinger theory of thermal transport~\cite{luttingerTheoryThermalTransport1964}, the  scale magnetic effect was translated into an unconventional thermoelectric transport coefficient in 3D untilted Dirac and WSMs~\cite{chernodubGenerationNernstCurrent2018, arjonaFingerprintsConformalAnomaly2019,arjonaromanoNovelThermoelectricElastic2019}. 
The very concrete prediction in Ref. \cite{chernodubGenerationNernstCurrent2018} of the thermoelectric coefficient dependence on the magnetic field and
temperature $\sim B/T$ is now being explored experimentally by several groups. For these delicate experiments, done as closed as possible to the conformal limit T=0 and zero chemical potential, it is crucial to keep control of all the variables. The tilt is an
important factor present in almost all real Dirac materials that does not break the scale invariance.

In this paper, we analyze the effect of the tilt in the previous context and develop a theory to investigate  new unconventional thermomagnetoelectric effects in type-I Dirac and WSMs. We show that a tilt-dependent transverse current is induced by a temperature gradient in the presence of a perpendicular magnetic field.  In tilted Dirac materials with full inversion symmetry, the tilt component parallel to the magnetic field gives rise to a novel axial (valley) current. Since the tilt  is a generic feature of the topological semi-metals, this study will be useful in the design and analysis of  experiments related to the thermoelectric transport  at very low temperatures and electron densities. 
\section{The model}
We consider an undoped topological WSM with and without inversion symmetry. 
Within the continuum model,  a tilted cone with chirality $s=\pm$ is described by the Hamiltonian \cite{Tilt2008},
\begin{equation}
     \mathcal{H}_s = s v_F {\sigma}^i  {p}_i + v_F {t}^i_s  {p}_i I_2 ,
\label{eq:Ham}     
 \end{equation}
where \( v_F \) is the isotropic Fermi velocity, $\sigma^{i}\!,\ i=1,2,3$ are the Pauli matrices,  \( \vec{p} \) is the momentum operator, and 
\( \vec{t}^s \) is the tilt vector. 
The eigen-energies of the Hamiltonian  are
\begin{equation}
E_{\lambda s}=\lambda v_F |\vec{p}|+v_F  {t}^i_s {p}_i, 
\end{equation}
where $\lambda=\pm$ refers to the  conduction and valence bands,
and we have set the reduced Planck constant $\hbar=1$.

For the inversion symmetric, case we have \( \vec{t}_s = s \vec{t} \), while for the broken inversion symmetry case, we will fix \( \vec{t}_s = \vec{t} \). The transition from type I to type II  WSMs occurs when  
$\vert t \vert >1$ \cite{tchoumakovMagneticFieldInducedRelativisticProperties2016}.
We will consider only type I tilted WSMs because the type II have an extended Fermi surface and cannot be modelled with the linear dispersion relation properly. Moreover the finite density of states at the Fermi level introduces a dimension-full parameter in the model that breaks the scale invariance explicitly. We will investigate the type II WSMs elsewhere.

In four-dimensional combined space-time notation, 
the fermionic Lagrangian density is given by \cite{armitageWeylDiracSemimetals2018}
\begin{equation}
   \mathcal{L}_{\text{F}}=i\bar{\psi}\left( \gamma^{0}\left( \partial_{0}-t^i\partial_{i}\right) +
   v_F\sum_{j=1}^3\gamma^{j}\partial_{j} \right) \psi  \  ,
   \label{LF}
\end{equation}
where $\psi$ is the fermionic field and $\gamma^{\mu}$ are the standard contravariant gamma matrices \cite{Peskin}.

Since the tilt vector coupling is dimensionless, the classical action remains scale invariant and   the corresponding \emph{canonical} energy-momentum tensor \cite{forgerCurrentsEnergyMomentumTensor2004},
\begin{align}
  \label{eq:energy-momentum-tensor}
  \mathcal{T}_s^{\mu\nu} =
  \frac{1}{2} (
  \Psi_s^{\dagger} \gamma_s^{\mu} \partial^{\nu} \Psi_s
  +\gamma_s^{\nu} \Psi_s^{\dagger} \partial^{\mu} \Psi_s
  -\eta^{\mu \nu} \mathcal{L}_s
  ),
\end{align}
is  traceless $(\mathcal{T}_s)^{\mu}_\mu=0$. We consider a flat space Minkowski metric $\eta^{\mu\nu}=\textrm{diag}(+1,-1,-1,-1)$.
In Ref. \cite{chernodubGenerationNernstCurrent2018} it was shown that, in  the untilted case, the conformal anomaly in the presence of electric and magnetic fields, together with the Luttinger effective gravitational
field theory of thermal transport, gives rise to an unconventional contribution to the thermoelectric transport coefficient. The formalism followed in Ref. \cite{chernodubGenerationNernstCurrent2018} was explicitly Lorentz covariant. Since the tilt violates rotational invariance, we cannot follow this approach and we will compute the expected unconventional contribution with a Kubo formalism akin to that used in Ref. \cite{arjonaFingerprintsConformalAnomaly2019}.

\section{The Luttinger approach to the thermoelectric transport coefficient.}
We will compute  the  thermoelectric transport coefficient in the presence of an applied magnetic field within a Kubo formula
as devised by Luttinger in Ref. \cite{luttingerTheoryThermalTransport1964}.

We chose the magnetic field to point along the $z$ direction, $\vec{B}=B \hat{\bm{z}}$ and include it in the Hamiltonian \eqref{eq:Ham}
by the minimal coupling $\vec{p} \rightarrow \vec{p} + e \vec{A}$
 where $e=|e|$ is the electron charge and $\vec{A}=-B y \hat{\bm{x}}$ is the  vector potential in the Landau gauge. 
 For each chiral Weyl spinor, the Hamiltonian in the presence of the magnetic field reads,

\begin{equation}
     \mathcal{H}_s = s v_F {\sigma}^i  ({p}_i+A_i) + v_F {t}^i_s  {p}_i I_2 ,
\label{eq:Ham2}     
 \end{equation}
We first assume that the tilt vector has a projected component along both the magnetic field $\vec{t}_s=t^z_s \hat{\bm{z}}$ and perpendicular to it $\vec{t}_s=t^x_s \hat{\bm{x}}$, and find the eigensystem of the Hamiltonian. Then, we investigate the effect of each component on the unconventional quantum transport in tilted undoped Weyl semimetals, separately.

To find the unconventional transverse thermoelectric response of undoped tilted WSMs in the presence of a temperature gradient, we use the Luttinger approach \cite{luttingerTheoryThermalTransport1964,tataraThermalVectorPotential2015,chernodubThermalTransportGeometry2021}. We introduce a gravitational scalar potential field $\psi$ that couples to the energy density $\mathcal{T}^{00}$, into our electronic Hamiltonian as a perturbation $\mathcal{H}_L=\int \mathrm{d}\vec{r} \psi \mathcal{T}^{00}$. The gradient of the gravitational field and temperature gradient $\bm{\nabla} T$ are related by $\bm{\nabla}\psi + {\bm{\nabla} T}/{T} = 0$. Within the Kubo linear response formalism, the response of the system to the gradient of gravitational field can be found by applying the conservation of the energy--momentum tensor: $\partial_{\mu}T^{\mu0}=\partial _t T^{00}/v_F  + \partial _i T^{i0} = 0$.  
Within this formalism, the transverse thermoelectric susceptibility $\chi_s^{ij}$ of tilted Weyl quasiparticles, defined by,
\begin{equation}
J_s^i=\chi_s^{ij}(-\partial_j T/T),
\label{eq:chi}
\end{equation}
is given by the following retarded response function \cite{arjonaFingerprintsConformalAnomaly2019}, 
\begin{equation}
  \label{eq:kubo-response}
  \begin{multlined}
    \chi_s^{ij}(\omega, \vec{q}) = \frac{-i v_F}{\mathcal{V}}
    \int \mathrm{d} t
    e^{i \omega t}
    \int\limits_{-\infty}^0 \mathrm{d} t'
    \Theta(t)\\
    \times
    \Braket{\left[J_s^i(t, \vec{q}), \mathcal{T}^{j0}(t', -\vec{q})\right]},
  \end{multlined}
\end{equation}
where $\mathcal{V}$ is the system volume and $\Theta$ is the Heaviside step function. 
The details of the computation  are depicted in the Appendix \ref{App}.

With all the previous ingredients (see Appendix \ref{App}, we find the static and long-wavelength unconventional transverse response function of an undoped tilted WSM in the limit \( T\to 0 \) to be,
\begin{equation}
  \label{eq:response-w-dimensions}
  \lim_{\omega \to 0} \lim_{\vec{q} \to 0}
  \chi_s^{xy} =
  \gamma_{N}(\bm{t}_s) \frac{e^2 v_F B}{2\hbar(2\pi)^2},
\end{equation}
where $ \gamma_N $ is a normalization factor that depends on the number of LLs $N$ that are taken into account in the evaluation of the response function, and more importantly, on the tilting vector \( \vec{t}^s \). This equation is our main result.

\section{The influence of the tilt on the thermoelectric coefficient}
In general, we can decompose the normalization factor $ \gamma_N $ in eq. \eqref{eq:response-w-dimensions} into an even and an odd function of the tilting parameter as $\gamma_N^{\text{even}(\text{odd})}(\bm{t}_s)=\big( {\gamma_N(\bm{t}_s) \pm \gamma_N(-\bm{t}_s)}\big)/{2}$.
In the following, we will compute the normalization factor $ \gamma_N $ for type-I  
WSMs in different tilting geometries.

For the sake of completeness, we first look at the untilted case, where was first introduced in Refs. 
\textcite{chernodubGenerationNernstCurrent2018} and \textcite{arjonaFingerprintsConformalAnomaly2019}.
We have simplified their result and found an analytical expression for $\gamma_N(\bm{t}^s=0)$ with $N\geq0$, as 
$$\gamma_{{N}} - \gamma_{{N}-1} = \gamma_{N=0} + 2 {N} \left\{ 1 - (1+{N}) \log \left(1 + {{N}^{-1}}\right) \right\},$$ where $\gamma_{N=0}=1$ and by definition $\gamma_{N=-1}=0$.
\\

\begin{figure}
  \centering
  \includegraphics[width=.50\textwidth]{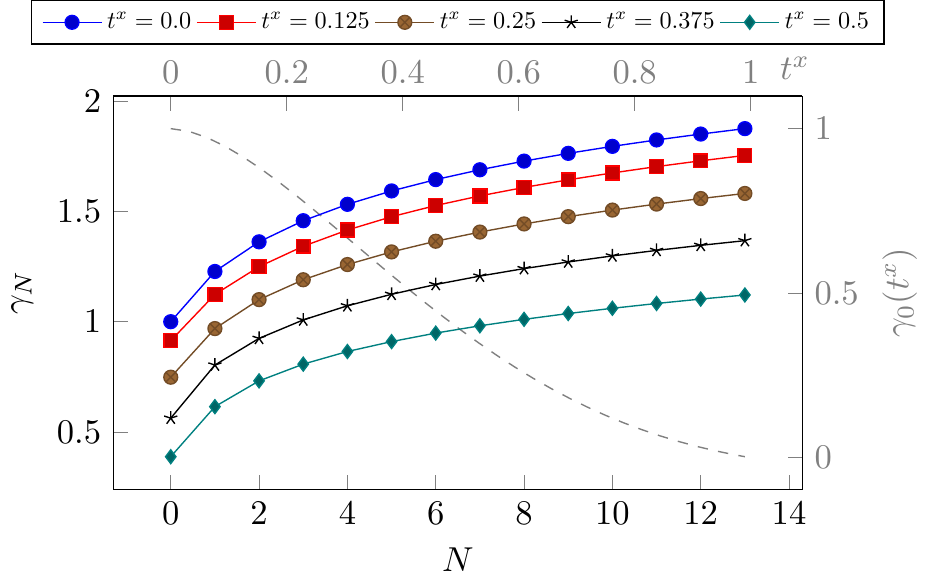}
  \caption{ The normalization factor $ \gamma_N $,  see Eqs. \eqref{eq:response-w-dimensions} and \eqref{eq:119}, as a function of the  LLs cutoff N (solid lines) for various values of the tilt parameter  perpendicular to the magnetic field direction \(\bm{t}= t_s^x \hat{\bm{x}}\). The plot also present the tilt-dependence of the normalization factor for lowest LLs cutoff $N=0$ (dashed gray line).}
  \label{fig:contribtx}
\end{figure}

\subsection{Tilt perpendicular to the applied magnetic field}
When the tilt vector of the Weyl cone is perpendicular to the applied magnetic field, $\bm{t}\perp\bm{B}$, we find,
 \begin{align}
    \label{eq:119}
    \gamma_{N}(t_s^x) =&
    -4 \hbar v_F \beta^3
    \sum\limits_{m>0,n\leq0}^{N}
    \int \mathrm{d} k_z
    e^{-P^2}\\ \nonumber \times
    &\frac{\alpha_{k_z m s}^2 (\Xi^{(1)}_{m,n, s})^2
      (E_{k_z m s} + E_{k_z n s})}{
      (\alpha_{k_z m s}^2 + 1)(\alpha_{k_z n s}^2 + 1)
      (E_{k_z m s} - E_{k_z ns})^2
    },
  \end{align}
where the integration limit is $ (-\infty, \infty) $, except for $ n = 0 $, where it is $ [0, \infty) $.
In this case, the normalization factor is an even function of tilting, and thus independent of the chirality $ s $.
The response was solved numerically and the result is shown in Fig. \ref{fig:contribtx}.
As the tilt increases, the $\gamma_{N=0}$ decreases monotonically. In Fig. \ref{fig:contribtx}, we present the normalization factor $\gamma_N$ as a function of the LL cutoff N, for various tilt parameters $t^x$.

\begin{figure}
  \centering
  \includegraphics[width=.9\columnwidth]{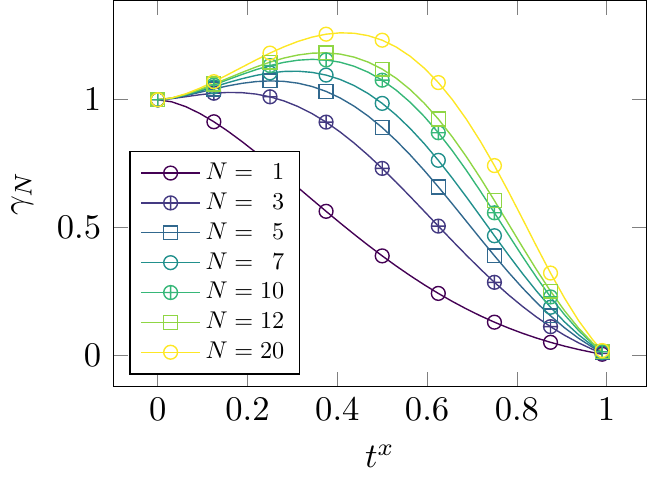}
  \caption{The normalization factor \( \gamma_{N} \), Eqs. \eqref{eq:response-w-dimensions} and \eqref{eq:119}, as a function of the tilt parameter perpendicular to the magnetic field direction \( \bm{t}= t_s^x \hat{\bm{x}}\), for various LLs cutoff $N$.}
    \label{fig:0tontx}
\end{figure}

The sum over LLs is bounded by considering all transitions between LLs below some cutoff level ${N}$.
In the untilted case, in the deep quantum limit where only the lowest LL is filled, one may only take into account \( 0\to \pm 1 \) transitions.
However, in the case of perpendicular tilt, where the dipolar selection rule \( |m|=|n|+1 \) is not valid anymore, it might be reasonable to consider higher transitions $ 0\to n $ with $|n| \leq {N} $.
Considering these transitions, we find a qualitatively different response. The response of the system shows a nonmonotonic behavior as a function of tilting parameter,  see Fig. \ref{fig:0tontx}. It has a maximum at a critical titling value and then decreases and vanishes as \( |t^x| \to 1 \). We will discuss   this emergent feature in Sec.
\ref{sec:discus}.

\subsection{ Tilt parallel to the magnetic field}
When the tilt vector is parallel to the magnetic field, it can be shown that  
the response of the tilt term in the Hamiltonian is algebraically added to the untilted response function, 
$\gamma_N(t_s^{z}) = \gamma_{N}(\bm{t}^s=0) + \delta\gamma_{N}(t_s^{z})$, where,
\begin{align}
  \label{eq:5}
  \delta\gamma_{N}(t_s^{z}) =&  -8 t_s^{z} \hbar^2 v_F^2 \sum\limits_{i=0}^{N} \int^{k^c_z}_{-\infty} \mathrm{d}k_z k_z \frac{\alpha_{k_z m s}^2
  }{\left[ (\alpha_{k_z ms}^2 + 1) (\alpha_{k_z ns}^2 + 1) \right]}\nonumber\\
  &\times \frac{n_{k_z ms} - n_{k_z ns}}{
   (E_{k_z m s} - E_{k_z n s})^2
  }\Big|_{\overset{m=i+1}{n=-i}},
\end{align}
with $n_{k_z ms}$ is the equilibrium Fermi-Dirac distribution function. Since we are using a linearized model Hamiltonian to describe the low-energy Weyl quasiparticles, this contribution is logarithmically divergent, so we introduce an ultraviolet wavenumber cutoff $k^c_z$ to get,
\small
\begin{align}
    \label{eq:6}
    &\frac{\delta\gamma_{N} - \delta\gamma_{N-1}}{ 2 t^{z}_s} =
    \Lambda \left(\sqrt{\Lambda^2 +N} -\sqrt{\Lambda^2 + N + 1} \right)
    \nonumber \\&+ (N+1) \tanh^{-1} \left[\frac{\Lambda}{\sqrt{\Lambda^2 + N+1} }\right]
    - N \tanh^{-1} \left[ \frac{\Lambda}{\sqrt{\Lambda^2 + N }} \right],
\end{align}
\normalsize
for $N\geq 0$, where $\Lambda= l_B k^c_z/\sqrt{2}$ is the dimensionless wavenumber cutoff and by definition $\delta\gamma_{-1}=0$. 
We estimate the ultraviolet cutoff as $k^c_z \approx 1/a$, where $a$ is the lattice constant of a Weyl or Dirac semimetal.

\begin{figure}[h]
  \centering
    \includegraphics[width=.45\textwidth]{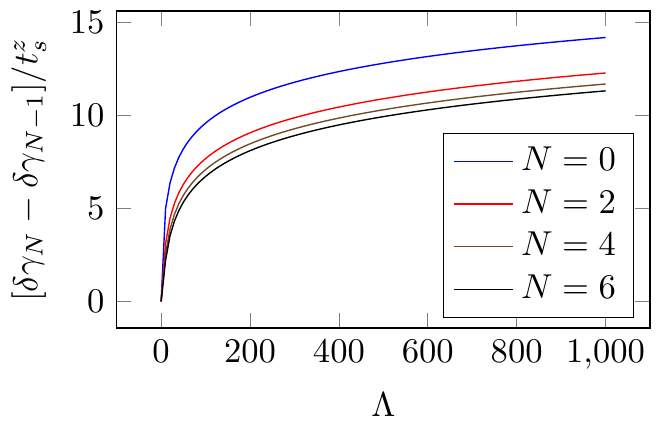}
\caption{The normalization factor $\gamma_N$, Eqs. \eqref{eq:response-w-dimensions} and \eqref{eq:6}, as a function of the dimensionless wavenumber cutoff for various LLs cutoff $N$, when the tilt parameter ${\bm{t}}=t_s^z \hat{\bm{z}}$ is parallel to the magnetic field direction.} 
 \label{fig:tz}
\end{figure}
Figure \ref{fig:tz} shows the contribution to the total normalization factor of a finite tilt parameter parallel to the magnetic field,  as a function of the ultraviolet wavenumber cutoff and for various LLs cutoff. 

This term is odd in the tilt vector, and thus in the inversion symmetric case, its contribution to the total normalization factor of the two Weyl cones is zero, and thus the tilting has no contribution to the total charge current ${J}_c^i$. However, this term will have a contribution to a valley (or axial) current $ {J}_A^i\equiv J^i_+-J^i_-$. Therefore, we will find an ``axial Nernst" 
current perpendicular to an applied magnetic field and to a gradient of temperature,  in the conformal limit (zero temperature and zero chemical potential)  in inversion symmetric tilted WSMs. 

\section{Discussion}
\label{sec:discus}
In this section we will discuss the salient features arising from Eq.(\ref{eq:119}). For any direction of the tilt vector $\bm{t}_s$, the function $\gamma_{N}(\bm{t}_s)$ is not bounded when including more Landau level transitions in the computation of Eq. (\ref{eq:119}). This response is also (logarithmically) divergent as the cutoff $\Lambda$ grows, as we can see in Fig. (\ref{fig:tz}) for the particular case $\bm{t}_s=t^z_s\hat{\bm{z}}$, in contrast to, for instance, what happens with the electric conductivity. This has  been discussed previously in untilted case \cite{arjonaFingerprintsConformalAnomaly2019}. The physical origin of this divergence is traced to the larger engineering dimension of the energy-momentum operator $\mathcal{T}^{0j}$ compared with the current operator $J^{i}$, as it can be readily seen in Eq.(\ref{eq:energy-momentum-tensor}). Also, as we are working in the limit of zero temperature and the Fermi level crossing the nodal point, transport properties are dominated by interband transitions, so we expect to obtain more reliable results when incorporating higher LL transitions, controlled by the cutoff parameter $N$.
What is more surprising is the presence of a maximum in $\gamma_{N}(\bm{t}_s)$, when the tilt vector is perpendicular to $\mathbf{B}$ (in our case, $\bm{t}_s=t^x\hat{\bm{x}}$), as it can be appreciated in Fig. (\ref{fig:0tontx}). In absence of tilt, $\gamma_{N}(\bm{t}_s=0)=1$, that agrees with previous results based on the conformal anomaly \cite{chernodubGenerationNernstCurrent2018}. However, as discussed in the introduction, the presence of a tilt vector $\bm{t}_s$, being a velocity, does not spoil scale invariance at the classical level as it does not introduce any scale in the problem, so a particular preferred value of $\bm{t}_s$ should not be expected. Also, the presence of a maximum in response functions often implies the competition of several effects. Besides, this maximum does not appear when $\bm{t}_s$ is parallel to the magnetic field $\mathbf{B}$. In fact, a linearly increasing $\gamma_N$ is expected (at least at small values of $\bm{t}_s$) as the current operator $J^{i}_s$ possesses a linear contribution with $\bm{t}_s$, as it can be seen in Eq. (\ref{eq:current-op}), for any tilt direction, and this dependence qualitatively explains why $\gamma_{N}$ grows when $\bm{t}_s=t^x_i\hat{\bm{x}}$ in Fig. \ref{fig:0tontx}. To understand the change of monotony in $\gamma_{N}(t^x_i\hat{\bm{x}})$ it is instructive to interpret the presence of the tilt vector in the Weyl Hamiltonian in terms of an emergent Lorentz symmetry \cite{udagawaFieldSelectiveAnomalyChiral2016,tchoumakovMagneticFieldInducedRelativisticProperties2016,yuPredictedUnusualMagnetoresponse2016,largeenhancement2019}. 
\begin{figure}
  \centering
  \includegraphics[width=.9\columnwidth]{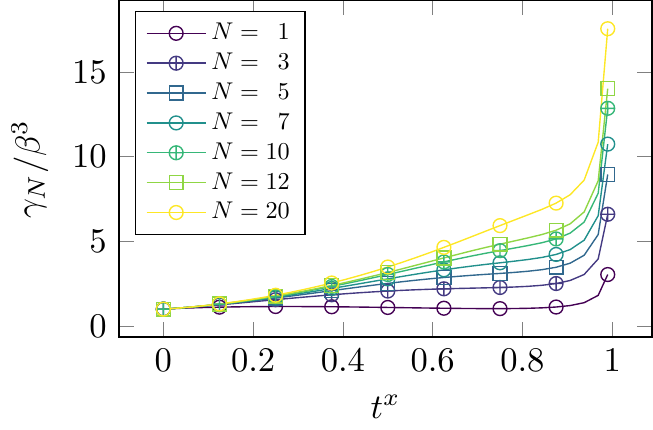}
  \caption{Same situation as Fig. \ref{fig:0tontx}  but with the normalization factor divided by $\beta^3$:  $\gamma_N/\beta^3$ as a function of the value of the tilt perpendicular to the magnetic field $\bm{t}_s=t^x_s\hat{\bm{x}}$.}
    \label{fig:normalizationfactorwithoutbeta}
\end{figure}
WSMs possess an emergent symmetry in form of Lorentz invariance, where the role of the speed of light is played by the Fermi velocity $v_F$, as can be readily seen in the Lagrangian $\eqref{LF}$. The presence of a tilt vector breaks this emergent Lorentz invariance selecting a preferred reference frame (lab frame) with velocity $\bm{t}_s$. Although this symmetry is broken by the tilt, we can still Lorentz transform the Hamiltonian to get rid of the tilt (tilt frame), compute the spectrum as in absence of the tilt and get back to the lab frame. The price to pay is that energies, momenta and other vectors get modified (rescaled) by the presence of the squeezing factor $\beta = \sqrt{1 - (\bm{t}_s)^2}$, which is the rapidity in the language of special relativity, and it is permissible when $\bm{t}_s<1$. We also know that in this regime, the effective magnetic field $\mathbf{B}$ acting in the tilt frame is reduced (squeezed) with respect to the magnetic field in the lab frame, $\mathbf{B}'=\beta^3 \mathbf{B}_{lab}$ \cite{yuPredictedUnusualMagnetoresponse2016}. This squeezing factor $\beta^3$ appears explicitly in Eq.(\ref{eq:119}) accompanying the magnetic field $\mathbf{B}(=\mathbf{B}_{lab})$. We have plotted in Fig. \ref{fig:normalizationfactorwithoutbeta} the normalization factor $\gamma_N$ divided by $\beta^3$. Now we observe that the rescaled normalization factor $\gamma_N/\beta^3$ monotonically grows with the tilt parameter, linearly for small tilt, due to the presence of the tilt term in the current operator $J^{i}_s$, and it diverges (within our numerical accuracy) when $t^x$ approaches to 1. This divergence appears in other transport properties of tilted WSMs \cite{tchoumakovMagneticFieldInducedRelativisticProperties2016,largeenhancement2019}. Then, the presence of a maximum in $\gamma_{N}$ for tilts perpendicular to the magnetic field is a competition from the fact that there is a piece of the electric current that is enhanced due to the tilt at low tilts (and eventually diverges when $t^x\to 1$), and the squeezing factor that modifies the effective magnetic field that decreases. 

\begin{figure}
  \centering
  \includegraphics[width=.9\columnwidth]{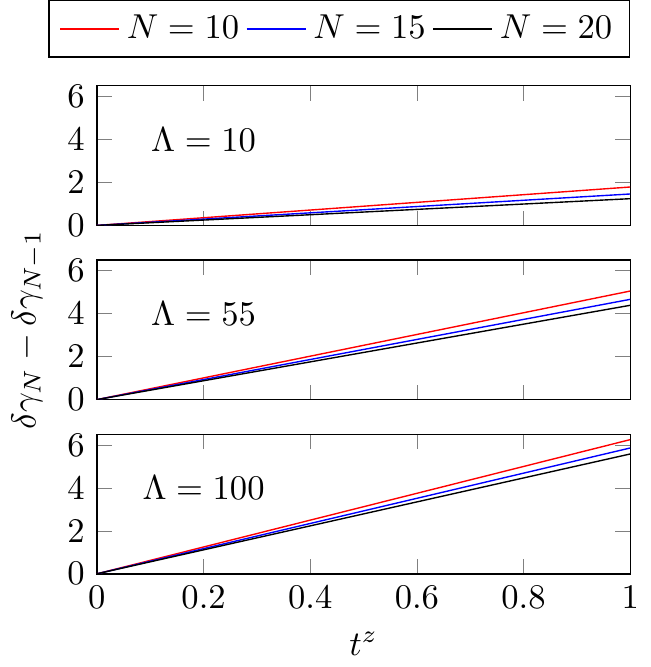}
  \caption{Linear behavior of the normalization factor as a function of the  tilt parameter parallel to the magnetic field, Eq. \eqref{eq:6}, for various values of the wavenumber cutoff $\Lambda$ and LL truncation $N$.}
    \label{fig:tiltparallel}
\end{figure}

When the tilt vector points along the magnetic field, such squeezing of the effective magnetic field does not occur, so the only term depending of the tilt vector is the one present in the current operator and we expect a monotonously increasing behavior (linear at low $t^z_s$) of $\gamma_N$. This behavior is plotted in Fig. \ref{fig:tiltparallel}.
Both effects are kinematic effects and do not spoil the scale anomaly responsible of the universal value at $\bm{t}_s=0$.  

To finish this section, we would like to comment on what we can expect for the transport coefficient $\chi^{xy}$ when the tilt, being perpendicular to the magnetic field, points along the direction $\hat{\bm{y}}$ (the direction of $ {\vec \nabla}T$) instead of pointing along $\hat{\bm{x}}$ (the direction of the computed current), as discussed in the previous section. According to the point of view adopted here, under the condition of the tilt vector $\bm{t}$ to be perpendicular to $\mathbf{B}$, a Lorentz transformation made to the tilted system is formally equivalent to study the system with no tilt but in the presence of effective magnetic and electric fields, $\mathbf{B}'$ and $\mathbf{E}'\sim \bm{t}$, the latter pointing along $\hat{\bm{y}}$. This implies the presence of a drift velocity that points perpendicularly to both $\mathbf{B}'$ and $\mathbf{E}'\sim\bm{t}$, that is, $\bm{v}_d$ pointing along $\hat{\bm{x}}$ that also scales with $t^y$ as it does $\mathbf{E}'$. This drift velocity would generate a component of the current $J^x$ that scales with $t^y$ that is not contained explicitly in the operator $J^{x}$ in Eq.(\ref{eq:current-op}), as, remember, the tilt now points along $\hat{\bm{y}}$. Also, as $\bm{t}$ is still perpendicular to $\mathbf{B}$, there will be a squeezing of the magnetic field in the transformed reference frame $\mathbf{B}'$. 

Then, we expect a similar qualitative behavior to the case when $\bm{t}=t\hat{\bm{x}}$ shown in Fig. \ref{fig:0tontx} and discussed at the beginning of this section. The coefficient of the thermoelectric conductivity will grow with $t$ for small values of $t^y$ due to the presence of the drift velocity along $\hat{\bm{x}}$ that scales linearly with $t^y$, and, for larger values of $t^y$ the squeezing of the effective magnetic field will dominate and make it decrease.

\section{Conclusion}

We have investigated the effect of Weyl cone tilting in Dirac and WSMs on the novel thermoelectric response based on the conformal anomaly, proposed in \cite{chernodubGenerationNernstCurrent2018,arjonaFingerprintsConformalAnomaly2019}. We show that the thermoelectric response function of undoped WSMs is sensitive to the tilt vector parameter and to its relative angle with respect to the external magnetic field. This dependence of the response function to the tilt parameter and magnetic field may be an important signature in experimental measurements to disentangle different thermoelectric responses in their signals and extract the conformal anomaly contribution. In particular, when the tilt vector is parallel to the magnetic field, the scale-anomaly argument still produces a lower bound to the corresponding Nernst coefficient, but this is not the case when the tilt vector is perpendicular to $\mathbf{B}$.
we  found an ``axial thermoelectric effect", i.e. the generation of an axial 
current perpendicular to an applied magnetic field and to a gradient of temperature, originated in the conformal anomaly 
in tilted WSMs with inversion symmetry when the tilt vector has is a projection in the direction of the magnetic field. It is interesting to mention that an ``axial" anomalous Nernst coefficient for tilted WSMs in the absence of external magnetic fields have been found to be proportional to the tilt vector, as the nodal separation is the vector playing the role of ``axial" magnetic field \cite{anomalousNernst}. The axial current as a response of the magnetic field, found in the present paper, is thus the reciprocal transport coefficient predicted in Ref. \cite{anomalousNernst}. 

The effects of the tilt of the cones on the transport properties of other Dirac materials in two and three spacial dimensions has been analyzed previously in the literature \cite{Cortijo18,Mandal20,Bergholtz20,Wang21,Mandal22}. The special feature of the present work lies on its connection to the conformal anomaly, which only occurs in (1+1) and (3+1) space-time dimensions.

\section*{Acknowledgment}
This project has been supported by the Norwegian Financial Mechanism Project No. 2019/34/H/ST3/00515, ``2Dtronics''; and partially by the Research Council of Norway through its Centres of Excellence funding scheme, Project No. 262633, ``QuSpin''. This work is also part of the Project No.  PGC2018-099199-B-I00 funded by Grant No. MCIN/ AEI/10.13039/501100011033/ and by ERDF A way of making Europe.
A.C. acknowledges financial support from the Ministerio de Ciencia e Innovación through Grant No.  PID2021-127240NB-I00 and the Ramón y Cajal program through Grant No. RYC2018- 023938-I

\appendix

\section{Kubo formula for the hermoelectric transport coefficient}
\label{App}
In what follows, we will set the basic ingredients to
compute the thermoelectric transport coefficient in the
presence of an applied magnetic field within a Kubo formula.
The Landau levels (LLs) of the tilted WSMs  have been computed in  \cite{yuPredictedUnusualMagnetoresponse2016,tchoumakovMagneticFieldInducedRelativisticProperties2016}.
As it is known, the 3D LLs are dispersing in the direction of the applied magnetic field. For the assumed 
geometry with the magnetic field pointing in the z direction, and using a generic plane wave ansatz  $\phi(\bm{r})=e^{ik_xx+ik_zz} \phi(y)$, the spectrum of \eqref{eq:Ham2}
is given by 
\begin{equation}
  \label{eq:eigenlevels}
  E_{k_z m s} =
  \begin{cases}
    t^{z}_s v_F k_z + \sign(m) v_F \beta \sqrt{2 e B \beta |m| + k_z^2}, & m \neq 0\\
    t^{z}_s v_F k_z - s \beta v_F k_z; & m = 0
  \end{cases}
\end{equation}
where $\beta$ is a \emph{squeezing factor} \( \beta = \sqrt{1 - (t_s^{x})^2} \). In absence of the tilt vector, $\bm{t}_s=0$ and $\beta=1$, this LL spectrum reduces to the dispersion of the untilted case $E_{k_z m s}^0(B)$ \cite{arjonaFingerprintsConformalAnomaly2019}.
The corresponding eigenstates are,
\begin{equation}
  \label{eq:eigenstates}
  \phi_m(y) = \frac{\beta^{1/2}   e^{\sigma_x \theta_s / 2 } e^{- \chi_m^2/2} }{{(L_{x} L_{z})}^{1/2}} 
  \begin{pmatrix}
    a_{k_z m s} H_{|m|-1} (\chi_m)\\
    b_{k_z m s} H_{|m|} (\chi_m)
  \end{pmatrix},
\end{equation}
where $H_n(x)$ are the Hermite polynomials, $\theta_s = - s \tanh^{-1} t_{x}$, $l_B=\sqrt{1/(eB)}$ is the magnetic length, and ${L_{x(z)}}$ is the system size along the $x(z)$ direction.  We also define

\( \chi_m = {\beta}^{1/2} (y/l_{B}-k_{x}l_B) + \frac{t^{x}_s l_B}{{\beta}^{1/2} v_{F}} E_{k_z m s}^0(\beta B) \), 
$$a_{k_z m s} =
                    \frac{
                    \alpha_{k_z m s} ({\beta/\pi})^{1/4}
                    }{
                    \sqrt{(\alpha^2 _{k_z m s} + 1)l_B2^{|m|-1} (|m|-1)!}
                    },$$ and 
$$ b_{k_z m s} =
                    \frac{
                     ({\beta/\pi})^{1/4}
                    }{
                    \sqrt{(\alpha^2 _{k_z m s} + 1)l_B 2^{|m|} |m|!}
                    };$$ 
with $$\alpha_{k_z m s} =\frac{-{(\beta |m|)}^{1/2}}{ \sign(m) s\sqrt{\beta |m| + l_B^2k_z^2/2} - l_B k_z/\sqrt{2}}.$$

It is interesting to note that  the LLs  collapse when $|t_s^{x}| \geq 1$ \cite{lukoseNovelElectricField2007,tchoumakovMagneticFieldInducedRelativisticProperties2016,arjonaCollapseLandauLevels2017}
which is  the condition for the transition from type I to type II WSMs. This can be seen from the expression of  $\beta$ and $\theta_s$.  Since the distance between the LLs is reduced by a factor $\beta^{3/2}$, the separation between LLs vanishes in the limit of  $|t_s^{x}| \rightarrow 1$ and $\beta\rightarrow 0$. On the contrary, there is no restriction on the value of the parallel tilt $t_s^{z}$. The energy levels are tilted by the parallel component of the tilt vector onto the magnetic field $t_{z}$, while the perpendicular component \( t_{x} \) squeezes the separation between the levels.

We will consider the charge current operator at each Weyl node, 
\begin{equation}
  \label{eq:current-op}
  {J}_s^i = s e v_F \tilde{\sigma}_s^i,
\end{equation}
where we have defined modified Pauli matrices as
\( \tilde{\sigma}_s^{i} = \sigma^{i} + s t_s^{i} \sigma^0 \). We can define the total charge current as ${J}_c^i =  {J}_+^i+{J}_-^i$ and charge-neutral valley current as $ {J}_v^i =  {J}_+^i-{J}_-^i$.
Without loss of generality, in the following, we assume $\bm{\nabla} T \parallel \hat{\bm{y}}$, and compute the transverse charge current $\vec{J}_s \parallel \hat{\bm{x}}$ in the presence of a magnetic field along the $z$ direction. 

In systems with broken time-reversal symmetry, the total current, computed within the linear response theory, is the sum of an observable transport current and a circulating orbital magnetization current that does not contribute to the observable current 
\cite{vanderwurffMagnetovorticalThermoelectricTransport2019,chernodubThermalTransportGeometry2021}. 
Since we are interested in the undoped case,  the diamagnetic-like orbital magnetization is zero and the Kubo formalism only gives the observable charge current \cite{vanderwurffMagnetovorticalThermoelectricTransport2019}.

To investigate the unconventional transport arising from the scale anomaly, we should compute the static transverse response of the system at the long-wavelength limit. After some tedious but straightforward analytical calculations \cite{Thorvald}, we find the charge current density $$J^x_s= s e v_{F} \beta^2 \sum_{\vec{k} m n}  J^x_{\vec{k} m s; \vec{k} n s}$$ and energy-momentum tensor $$T_{s}^{y0}=\frac{is \beta}{2}\sum_{\vec{k} m n}T_{\vec{k} n s, \vec{k} m s}^{y0}$$ expressed in terms of the corresponding matrix elements,  that are given by 
\begin{align}
    \label{eq:112}
  J^x_{\vec{k} m s; \vec{k} n s} =&                                   [ \alpha_{k_z m s} \Xi_{m,n,s}^{(1)}                    + \alpha_{k_z n s} \Xi_{m,n,s}^{(2)} ]\Gamma_{\vec{k} m n s},
 \\ T_{\vec{k}  n s, \vec{k} m s}^{y0}  =& 
     [\alpha_{k_z m s} \Xi_{m,n,s}^{(1)} - \alpha_{k_z n s} \Xi_{m,n,s}^{(2)}]\Gamma_{\vec{k}  m n s}\nonumber\\ &\times  (E_{k_z m s} + E_{k_z n  s}).
\end{align}
In the above equations, we have defined 
\begin{align}
 &\Gamma _{\vec{k} m n s} =  \nonumber \\
&\frac{\exp\big[-({l_{B}^2(t^x)^2}/{4v^2_F\beta}) \big(E^0_{k_z n s}(\beta B)-E^0_{k_z m s} (\beta B)\big)^2\big]}{\sqrt{(\alpha_{k_z m s}^2 +1) (\alpha_{k_z ns} ^2 + 1)}}, \nonumber
\end{align}
and
\begin{subequations}
\begin{align}
\Xi^{(1)} &=
                            \begin{cases}
                                \Xi\big[|n|,|m|,P\big], & |n| \geq |m|-1\\
                                \Xi\big[|n|\leftrightarrow(|m|-1), P\rightarrow -P\big]; &  |n| \leq |m|-1
                              \end{cases}\nonumber\\ \nonumber
\Xi^{(2)} &=
                            \begin{cases}
                                \Xi \big[|n|\leftrightarrow |n-1|, |m|\leftrightarrow |m-1|\big], &  |n|-1 \geq |m|\\ \nonumber
                                \Xi\big[|n|\leftrightarrow |m|,P\rightarrow -P\big]; &  |n|-1 \leq |m|
                              \end{cases}
\end{align}
\end{subequations}
where $\Xi = \sqrt{\frac{2^{|n|} (|m|-1)!}{2^{|m|-1} |n|!}}                                     \left( \frac{P}{\sqrt{2}} \right)^{|n|-|m| + 1}                             L^{(|n|-|m|+1)}_{|m|-1} (P^2)$,
with $L_n^{(\alpha)}(x)$ is the generalized Laguerre polynomial and $P= {s t^x l_B \big( E^0_{k_z n s}(\beta B) - E^0_{k_z m s}(\beta B) \big)}/(v_{F} \sqrt{2 \beta })$.

\bibliography{Refs.bib}
\end{document}